\documentclass[12pt,fleqn,twoside]{article}
\usepackage{amsfonts}
\usepackage{espcrc1}

\input{tcilatex}

\begin{document}

\title{\textbf{Modelling gap-size distribution of parked cars using
random-matrix theory}}
\author{A.Y. Abul-Magd \\
Department of Mathematics, Faculty of Science, Zagazig University, Zagazig,
Egypt}
\date{\today}
\maketitle

\begin{abstract}
We apply the random-matrix theory to the car-parking problem. For this
purpose, we adopt a Coulomb gas model that associates the coordinates of the
gas particles with the eigenvalues of a random matrix. The nature of
interaction between the particles is consistent with the tendency of the
drivers to park their cars near to each other and in the same time keep a
distance sufficient for manoeuvring. We show that the recently measured
gap-size distribution of parked cars in a number of roads in central London
is well represented by the spacing distribution of a Gaussian unitary
ensemble.


\end{abstract}

\section{Introduction}

Random-matrix theory (RMT) \cite{mehta} models quantal chaotic systems in
terms of three canonical ensembles of random matrices. The Gaussian
orthogonal and symplectic ensembles (GOE and GSE) are used to describe
time-reversal invariant systems with integer and half-integer spins,
respectively, while the Gaussian unitary ensemble (GUE) is connected with
systems without time reversibility. The theory was originally proposed to
describe the fluctuation properties of energy spectra of atomic nuclei \cite%
{porter}. Bohigas et al \cite{bohigas} have established that the fluctuation
properties of the canonical random matrices are generic and therefore
applicable for the spectral statistics of a wide variety of quantal systems.
The broad range of applicability of RMT has recently been the subject of
many excellent reviews, e.g. \cite{guhr,mirlin}.

RMT is also applicable for classical one-dimensional interacting many
particle systems. Long time ago, Dyson \cite{dyson} realized that
eigenvalues of Gaussian random matrices do indeed behave like charges,
repelling each other with a force varying inversely with the first power of
distance. The matrix eigenvalues describe the positions of the charged
particles. He showed that the joint eigenvalue-distribution of Gaussian
matrices have exactly the same mathematical structure as the Boltzmann
factor for a one-dimensional classical Coulomb gas. The same holds true also
for other potentials. An example is the Pechukas gas \cite{pechukas}, where
one-dimensional particles interact by a potential inversely proportional to
their mutual distance. Recently, there have been numerous fascinating
applications of RMT outside quantum chaology. Representative examples are
the price fluctuations in a stock markets \cite{laloux,plerou}, the\
statistical properties of bus arrivals \cite{seba}, the atmospheric
correlations \cite{santhanam} and the statistics of geometrical resonances
in disordered binary networks \cite{gu}.

The present paper proposes an application of RMT to the analysis of recent
empirical data on the gap distribution of parked cars in the streets of
London \cite{rawal}. These authors could not reproduce their data using
different versions of the random parking model, in which a car can park at
any space greater than or equal to its length \cite%
{krapivsky,rodger1,rodger2}. They had to introduce several additional
assumptions in order to achieve qualitative agreement with the empirical
results. Alternatively, we apply Dyson's Coulomb gas model \cite{dyson},
which allows for using the results of RMT. Section 2 is a brief review of
the model with a special emphasis on its relevance to real car parking.
Section 3 shows a comparison between the empirical gap-size distribution
measured in Ref. \cite{rawal} and the level-spacing distribution for GUE.
The conclusion of this work is formulated in Section 4.

\section{Coulomb gas model}

The purpose of this section is to show that Dyson's one-dimensional gas
model is relevant to the car parking problem. The model considers a gas of $N
$ point charges with positions $x_{1},x_{2},...,x_{N}$ free to move on the
infinite straight line $-\infty <x<\infty $. The potential energy of the gas
is given by \cite{mehta}%
\begin{equation}
V=\frac{1}{2}\dsum\limits_{i}x_{i}^{2}-\dsum\limits_{i<j}\ln \left\vert
x_{i}-x_{j}\right\vert .
\end{equation}%
The first term represents a harmonic potential that attracts each charge
independently towards the coordinate origin. The second term represents an
electrostatic repulsion between each pair of charges. The logarithm function
comes in if we assume the universe to be two-dimensional. Nevertheless, the
methods of the statistical physics remain valid also for different shapes of
the particle-particle interaction. For example, Krb\'{a}lek and \v{S}eba 
\cite{seba}\ have numerically evaluated the equilibrium distributions of one
dimensional gas interacting via two body potentials\ $v_{ij}\approx
1/\left\vert x_{i}-x_{j}\right\vert ^{a}$ with $a<2$ and found that the
resulting distributions belong to the same class as the Dyson case. Now, let
the charged gas be in thermodynamic equilibrium. The probability density
function for the position of the charges is given by the Boltzmann factor%
\begin{equation}
P(x_{1},x_{2},...,x_{N})=Ce^{-\beta V},
\end{equation}%
where $C$ is a normalization constant, $\beta =1/kT$ and $k$ is the
Boltzmann constant. Substituting (2) into (1), we obtain \cite{mehta,porter}%
\begin{equation}
P(x_{1},x_{2},...,x_{N})=C\dprod\limits_{i<j}\left\vert
x_{i}-x_{j}\right\vert ^{\beta }\exp \left( -\frac{1}{2}\beta
\dsum\limits_{k}x_{k}^{2}\right) .
\end{equation}%
This is exactly the joint probability density function for the eigenvalues $%
x_{1},x_{2},...,x_{N}$ of matrices from a GOE, GUE or GSE if $\beta =1,2$ or
4, respectively. Thus thus the role of the inverse temperature in the
Coulomb gas model is played by the level-repulsion power of eigenvalues of
the random matrices, which is specified by the symmetry of the ensemble with
respect to time reversibility.

RMT has elaborate methods for calculating the spectral characteristics of
the three canonical ensembles. For example, Mehta \cite{mehta} expresses the
nearest-neighbor spacing distribution $P(s)$ as second derivatives of an
infinite product of the factors $\left[ 1-\lambda _{n}\left( s\right) \right]
$, where $\lambda _{n}\left( s\right) $ are the eigenvalues of certain
integral equations. Here $s_{i}=\left( x_{i+1}-x_{i}\right) /D$ and $D$ is
the mean level separation and $s$ is a randomly chosen $s_{i}$. These
approaches result in tabulated numerical values, series expansions and
asymptotic expressions for the spacing distribution. They unfortunately do
not lead to closed-form expressions that can be in the analysis of
experimentally observed or numerically calculated discrete data.

In many cases, the empirical spacing distributions are compared to the
so-called Wigner surmise. The latter follows from the exact solution for $%
2\times 2$ matrices but still presents an accurate approximation to the
exact results obtained for ensembles of large matrices. The Wigner surmise
for GUE is given by%
\begin{equation}
P(s)=\frac{32}{\pi ^{2}}s^{2}e^{-\frac{4}{\pi }s^{2}}.
\end{equation}%
To demonstrate the accuracy of the Wigner surmise, we expand this
distribution in powers of $s$ to obtain%
\begin{equation}
P(s)=\frac{32}{\pi ^{2}}s^{2}\left( 1-\frac{4}{\pi }s^{2}+\cdots \right)
\cong 3.242s^{2}-4.128s^{4}+\cdots ,
\end{equation}%
while the power-series expansion of the corresponding exact distribution 
\cite{mehta} yields%
\begin{equation}
P_{\text{exact}}(s)=\frac{\pi ^{2}}{3}s^{2}-\frac{2\pi ^{4}}{45}s^{2}+\cdots
\cong 3.290s^{2}-4.329s^{4}+\cdots .
\end{equation}

The question which remains to be answered is whether the Coulomb model is
suitable for the car-parking problem. We see no problem in representing cars
of different size by point particles since we are eventually interested in
the gap-size distribution. We even dare to consider this as an advantage
because the results do not involve the (average) car length as in other
formalisms. The gap distribution, when represented by Eq. (4), has no
adjustable parameters once the gaps $s$ are measured in units of their mean
size $D$. Furthermore, we represent the cars by classical
electrically-charged particles in the state of thermodynamic equilibrium
using the Boltzmann factor (2).This factor is obtained from the
Gibbs-Boltzmann canonical distribution by integration over the momenta of
the particles. Namely this kind of averaging over the momenta is what allows
to link the stationary parked cars with the moving gas particles. The
particles are under the influence of a confining potential, which reflects
the preference of the drivers to park their cars close to the target of the
journey. A potential consisting of a sum of a single- and two-body terms, as
in Eq. (1), seems to provide a reasonable representation for the car parking
process. The single-particle term reflects the general tendency of the
drivers to park their cars not too far from each other. The repulsive
two-body term is meant as an expression of the tendency of the driver to
keep a distance from the other parked cars in order to allow for parking
manoeuvring. Thus, the superposition of these two potential, which is
repulsive for small gaps and attractive for the large ones, expresses the
fact that it is unlikely to see too small or too large gaps between parked
car. The proposed model may thus be viewed as a generalization of the random
car parking model, in which the simple geometric exclusion is replaced by
dynamical coupling through a repulsive potential. The correspondence between
RMT and the Coulomb-gas model is technical and based on identification of
the inverse temperature $\beta $ of the gas with degree of level repulsion
as mentioned above. The car parking problem has no time reversal symmetry.
For this reason, we assume GUE which violates time reversibility is more
suitable for the car parking problem than the other ensembles, and set $%
\beta =2$.

\section{Comparison with empirical results}

Rawal and Rodger \cite{rawal} measured the gaps between parked cars on four
connected streets without any driveway or side streets in central London.
The data were collected in the late evening so that there were few places
capable of taking additional cars. In total, 500 gaps were measured. The
average gap size was 154.2 cm. Figure 1 shows by a histogram the probability
distribution function $P(s)$, normalized to unity, for the gaps measured by
Rawal and Rogers \cite{rawal}. The spacings $s$ are defined as the ratio of
the gaps to their mean value.  The normalization is done in such a way that
the overall surface under the data is 1. These modifications of the original
data are introduced in order to make the comparison with the
nearest-neighbor spacing distribution for GOE, which is represented in Fig.
1 by a solid curve. We can go further by regarding the renormalized data as
a probability distribution for gaps in an ensemble of equivalent streets if
the streets involved in the experiment by Rawal and Rodger do not repesent a
special case. In the opposite case, the results of this experiment would be
of low value. We can then assign a statistical error for each gap interval
equal to $\sqrt{n_{i}},$where $n_{i}$ is the number of car pairs separated
by the corresponding gap. The resulting statistical uncertainties are shown
by error bars.

The empirical gap-size distribution \cite{rawal} does not agree with the
random car parking model. In this model, a particle of length $l$ is
randomly deposited along a linear chain if it finds an empty place of length
greater than or equal to $l$. The frequency of small gaps obtained in this
model is extremely high and monotonously decreases as the gap size increases 
\cite{krapivsky}. The empirical distribution behaves in a different way. The
frequency of small gaps are small and increases as the gap size increases
until it reaches a maximum at a gap size slightly less than the average
size. In order to fit the empirical distribution Rawal and Rodger proposed
two generalizations of the car-parking model, which were called model A and
model B. In model A, cars are allowed to park only if the space is larger
than $l+\varepsilon $, where the extra space $\varepsilon $ give a room to
manoeuvre. They obtained a distribution that takes a nearly constant value
at small gaps and then decreases after the gap size reaches a value of
approximately $0.2l$. Their model B suggests that car parking is governed by
two competing mechanisms. One is from people who just park anywhere and the
other from those who maneuver to make better use of the available space.
Thus, in each time step, a car is parked in an empty space and remains with
probability $p$, or further derives, with probability $1-p,$ to a distance $y
$ to the nearest car with a probability given by an arbitrary function $f(y)$%
. Model B provides a reasonable description of the data for the choice $p=0.3
$ and $f(y)=6y(l-y)$, which leads to the dashed curve in Fig. 1. However,
the disagreement at small gap sizes is still statistically significant.

We here show that the empirical gap size distribution of parked cars agrees
with the prediction of the Coulomb gas model as formulated in the previous
Section, without any additional assumptions. As mentioned above, GUE is
meant to describe systems which are chaotic and violate time reversibility.
The car parking process has both of these two features. The chaotic behavior
is caused by the human behavior of the drivers. The absence of
reversibility. is an essential feature of all random adsorption processes.
The solid curve in Fig. 1 is calculated using Eq. (4) for the spacing
distribution for a GUE. The calculation is done without any adjustable
parameter. The agreement between the predictions of RMT and the empirical
data is excellent. In comparison with traditional approaches to the car
parking problem, the Coulomb gas model includes fewer gaps with sizes
tending to zero. This suggests that "level repulsion" is an important
mechanism overlooked in the tradition car-parking process, and RMT may be
useful in understanding this process.

\section{Discussion}

We applied Dyson's Coulomb gas model to the car parking problem. This model
is relevant to real car parking as it represents the tendency of the driver
to keep a distance from the nearest parked car by repulsion between gas
particles and the tendency of the divers to park their cars close to each
other by the single-particle harmonic-oscillator potential. We analyzed the
empirical data on the distribution of the relative size of gaps between
parked cars, measured by Rawal and Rodger. We found an excellent agreement
between the empirical data and the spacing distribution for Gaussian unitary
ensemble of random matrices.

The significance of the present results is that they suggest to add the
random car parking problem to the long list of systems with RMT-like
fluctuations. The latter is the one-dimensional continuum version of the
random sequential adsorption process, in which particles land successively
and randomly on a surface. In this model, if an incoming particle overlaps a
previously deposited one, it is rejected due to the geometrical exclusion
effect. Without additional assumptions, the random sequential process cannot
explain gap size distribution as shown by Rawal and Rodger. In this respect,
the Coulomb gas model has the advantage of extending the repulsion between
parked cars further than their geometrical size. The random sequential
adsorption model enjoys a wide interest in physics, chemistry, biology and
in many other branches of science and technology \cite{evans,talbot}.
Obviously, the proposed model is no substitute for the elaborate
investigations of the random sequential absorption process. Nevertheless,
the powerful methods of RMT may be useful in understanding some aspects of
this important process.

\bigskip

\pagebreak

\bigskip \FRAME{ftbpF}{4.0266in}{3.442in}{0in}{}{}{Figure}{\special{language
"Scientific Word";type "GRAPHIC";maintain-aspect-ratio TRUE;display
"USEDEF";valid_file "T";width 4.0266in;height 3.442in;depth
0in;original-width 3.9781in;original-height 3.3952in;cropleft "0";croptop
"1";cropright "1";cropbottom "0";tempfilename
'IIC4AP00.wmf';tempfile-properties "XPR";}}

Fig. 1. The gap size distribution for parked cars, measured by Rawal and
Rodgers \cite{rawal} compared with the spacing distribution for GUE (solid
curve). The dashed curve is calculated using model B of Ref. \cite{rawal}

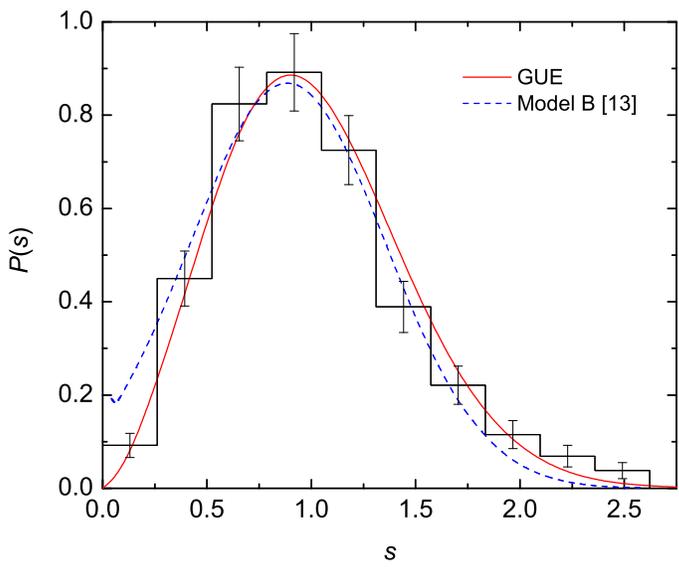
\end{document}